\documentstyle[12pt]{article}
\begin{document}
{\large

\begin{center}
{\bf \Large Polar Jahn-Teller centers and \\ 
magnetic neutron scattering cross-section in copper oxides.}
\end{center}

\begin{center}
A.S. Moskvin$^{\star}$, A.S. Ovchinnikov
\end{center}

\begin{center}
$^{\star}$Department of Theoretical Physics, Ural State University, 620083, Ekaterinburg, Russia 
\end{center}

 In the framework of the model of the polar singlet-triplet Jahn-Teller centers the 
cross-section is obtained for magnetic neutron scattering in high-$T_{c}$ cuprates.  Multi-mode character of the $CuO_{4}$ cluster ground manifold  in the new phase of polar centers determines the dependence of magnetic form-factor on the local structure and charge state of the center. It is shown that   magnetic inelastic neutron scattering in the system of the polar singlet-triplet Jahn-Teller centers permits to investigate the non-magnetic charge and structure excitations.

Corresponding author: Prof. A.S. Moskvin,  Department of Physics,   Ural State University, 620083, Ekaterinburg, Russia 

 Fax: +7-3432-615-978 	\ \ \ \ \  E-mail: alexandr.moskvin@usu.ru



\newpage{}

\section{Introduction.}

	Unconventional properties of the oxides like $YBa_{2}Cu_{3}O_{6+x}$,
 $La_{2-x}Sr_{x}CuO_{4}$, $(K,Ba)BiO_{3}$, $La_{1-x}Sr_{x}MnO_{3}$, $La_{2}CuO_{4+d}$, $La_{2}NiO_{4+d}$ including systems with the high-$T_{c}$ superconductivity and colossal magnetoresistance  reflect a result of the response of the system to the nonisovalent substitution that stabilizes phases providing the most effective screening of the charge inhomogeneity. These phases in oxides may involve novel unconventional molecular cluster configurations like the Jahn-Teller  center \cite{Moskvin} with anomalous high local polarizability and multi-mode behavior.	
	The numerous experimental investigations show that the origin of the high-$T_{c}$ superconductivity and other unconventional physical properties of the copper oxides is connected with an active interplay of the whole set of the degrees of freedom including the charge, spin, orbital and structural modes under conditions of the strong charge inhomogeneity and granularity.
	Indirectly this is corroborated by the failures to explain more or less completely the unconventional behavior of copper oxides as homogeneous systems within the "single-mode" approaches such as the spin-fluctuation, electron-phonon or purely electronic ones.
	An active interplay of the whole set of the modes is a natural element of the "multi-mode" scenario based on the so called "polar Jahn-Teller center model" proposed by  one of the authors earlier \cite{Moskvin, Loshka}. The $CuO_{4}$-clusters based copper oxides in this model are considered as systems of the local singlet bosons moving on the lattice composed from the singlet-triplet pseudo-Jahn-Teller (pseudo-JT) centers. An essential physics of cuprates within this scenario is connected with the multi-mode behavior and phase separation. 
	An occurrence of the unconventional properties of the copper-oxygen $CuO_{4}$ clusters as the basic elements of crystalline and electronic structure of the cuprates results in revision of many standard approaches. Below, we'll consider some features of the  magnetic neutron scattering  in  the system of the polar pseudo-JT centers and obtain an expression for the appropriate cross-section.

 It is worth to note that namely the magnetic inelastic neutron scattering together with the NMR and NQR stimulated the elaboration of the well known model of spin fluctuations for the cuprates \cite{Pines1, Pines2} which is considered as the most promising and perspective for the HTSC mystery explanation. We'll show that the developed model reveals some new features of the magnetic inelastic neutron scattering in cuprates.

\section{Polar singlet-triplet Jahn-Teller center model.}

The $CuO_{4}$-clusters based copper oxides within the polar pseudo-JT centers model  are considered as systems unstable with respect to the disproportionation reaction
 $$
	[CuO_{4}]^{6-} + [CuO_{4}]^{6-} \to [CuO_{4}]^{5-}_{PJT} + [CuO_{4}]^{7-}_{PJT}
$$       
with the creation of the system of the polar (hole  $[CuO_{4}]^{5-}$  or electron $[CuO_{4}]^{7-}$) Jahn-Teller centers. These centers are distinguished by the so called S-boson or two electrons paired in the completely filled molecular orbital of the $CuO_{4}$-cluster. In other words, the new phase can be considered as a system of the local spinless bosons moving in the lattice of the hole JT centers or the generalized quantum lattice bose-gas  with the boson concentration near $N_{B}=\frac{1}{2}$. The Jahn-Teller structure of the polar centers provides the high stability of the disproportionated phase with the small probability of the recombination process.
	A near degeneracy within ($^{1}A_{1g}$, $^{1,3}E_{u}$ )-manifold can create conditions for the anomalous strong electron-lattice correlations (pseudo-Jahn-Teller effect \cite{bersuker} ) with the active local displacement modes of the  $Q_{e_{u}}$, $Q_{b_{1g}}$ and $Q_{b_{2g}}$ types.
	In general, the pseudo-Jahn-Teller effect in the ($^{1}A_{1g}$, $^{1,3}E_{u}$)-manifold results in the formation of the four-well adiabatic potential of two symmetry types:  $E_{u}B_{1g}$ and $E_{u}B_{2g}$. In the first case we have four minima with the nonzero local displacements  $Q_{e_{u}} \not= 0$, $Q_{b_{1g}} \not= 0$ for the hybrid copper-oxygen mode $Q_{b_{1g}}$ and the purely oxygen mode $Q_{e_{u}}$. In the second case four wells correspond to the nonzero displacements of the $Q_{e_{u}}$  and $Q_{b_{2g}}$-types. 
          A type ($E_{u}B_{1g}$ or $E_{u}B_{2g}$) of the ground JT mode has the principal importance for  the physics of the copper oxides. It is determined by the competition between the vibronic parameters for the $Cu3d-O2p$ and $O2p-O2p$-bonds minimizing the $E_{u}B_{1g}$  and $E_{u}B_{2g}$-modes, respectively. In the real systems, we deal with a ground "rhombic" $E_{u}B_{1g}$ -mode \cite{panov}. Fig. 1 gives the qualitative picture for the pseudo-JT effect in the hole  $[CuO_{4}]^{5-}$ cluster with  illustrations of the cluster deformations in the $E_{u}B_{1g}$ and $E_{u}B_{2g}$ modes. In both cases we have to deal with the ground state JT quartets which undergoes the tunnel splitting to one doublet and two singlets. 

The hole pseudo-JT center with its high polarizability can be a center of an effective local pairing with a formation of the local singlet boson or  two electrons paired in completely filled molecular shell. The hole $[CuO_{4}]^{5-}_{PJT}$ center with the local boson represents the unconventional electron center  which is essentially distinguished from the "primitive"  $[CuO_{4}]^{7-}$ center considered as a non-degenerate system with the completely filled $Cu3d$  and $O2p$-shells.
	A transfer from the hole to the electron pseudo-JT center because of charge fluctuations is accompanied generally by the change of the local bare parameters such as A-E-separations $\Delta_{AE}=\varepsilon (^{1}E_{u}) - \varepsilon (^{1}A_{1g})$, singlet-triplet separation $\Delta_{ST}=\varepsilon (^{3}E_{u}) - \varepsilon (^{1}E_{u})$  and also by the change  of the ground JT mode ($E_{u}B_{1g} \leftrightarrow E_{u}B_{2g}$). In other words, the charge fluctuations in the phase of the pseudo-JT centers are strongly coupled with the local spin and structural fluctuations that results in complicated multimode behavior.

\section{The magnetic neutron cross-section in pseudo-JT center system.}

The neutron cross-section in the system of the pseudo-JT centers is determined by 
the common expression  \cite{izyumov}
$$
		\frac{d^{2}\sigma}{d \Omega d \varepsilon}=
\frac{1}{2 \pi \hbar} \frac{p'}{p}(r_{0}\gamma)^2 \sum_{\alpha, \beta = x,y,z} \frac{1}{3} (\delta_{\alpha \beta} - e_{\alpha}e_{\beta})\times
$$
\begin{equation}
\sum_{l,l'}\sum_{\nu,{\nu}'=1,2} \int^{+ \infty}_{- \infty} d t e^{-i {\omega} t} e^{-i{\vec q}{({\vec R}_{l'} -{\vec R}_{l})}}
\langle    
	s^{\beta}_{l'{\nu}'}e^{-i{\vec q}{\vec r}_{{\nu}'}}
	e^{i{\vec q}{\vec r}_{{\nu}}(t)} s^{\alpha}_{l{\nu}}(t)
\rangle,
\end{equation}
where sums run over all pseudo-JT centers $l, l'$ and over the holes of the separate pseudo-JT center $\nu, 
{\nu }'$; ${\vec s}_{l1}$, ${\vec s}_{l2}$ are the spins corresponding to the $b_{1g}$ and $e_{u}$ molecular orbitals of the $[CuO_{4}]^{5-}$ cluster two-hole configuration; $\vec e = \frac{\vec q}{q}$ is the unit scattering vector; $r_{0}$ is the electromagnetic electron radius;  $\gamma = -1.913$ is the neutron magnetic moment written in terms of Borh's magneton $\beta_{n}$.

Within the ${}^{3}E_{u}$ manifold the correlation function in (1) can be presented as
\begin{equation}
	\langle 
		S^{\beta}_{l'}F^{\star}(\vec q) F({\vec q}, t) S^{\alpha}_{l}(t)
	\rangle,
\end{equation}
where the total spin operator  ${\vec S}_{l} = {\vec s}_{1l}+{\vec s}_{2l}$ of 
the $[CuO]_{4}^{5-}$ cluster were used. 
Taking into account the vibronic interaction the form-factor of magnetic neutron scattering 
\begin{equation}
	F(\vec q, t)=\frac{1}{2}\sum_{\nu = 1,2} e^{i{\vec q}{\vec r}_{\nu}(t)}	
\end{equation}  
takes the operator form within the ${}^{3}E_{u}$ manifold,
e.g. for the $B_{1g}$-type adiabatic potential of trivial vibronic
$E+(b_{1g}+b_{2g})$ problem \cite{bersuker} (see Fig. 2) without account of the 
$pd$ and $pp$ overlap integrals 
\begin{equation}
	{\hat F}(\vec q, t)={F}_{0}(\vec q)+{F}_{1}(\vec q){\hat \sigma}^{z}(t),
\end{equation}
where $\sigma^{z}$ is Pauli's matrix,
$$
	{\hat F}_{0}(\vec q) = \frac{1}{2}|c_{d}|^{2} 
{\langle d_{x^{2}-y^{2}}| e^{i {\vec q}  {\vec r}}| d_{x^{2}-y^{2}}
 \rangle}+
$$
$$
\frac{|c_{p}|^{2}}{4}(\cos{(q_{x}R)} {\langle p_{x}| e^{i {\vec q}  {\vec r}}| p_{x}
 \rangle}+
\cos{(q_{y}R)} {\langle p_{y}| e^{i {\vec q}  {\vec r}}| p_{y}
 \rangle
})+
$$
$$
\frac{|c_{\sigma}|^{2}}{4}(\cos{(q_{x}R)} {\langle p_{x}| e^{i {\vec q}  {\vec r}}| p_{x}
 \rangle}+
\cos{(q_{y}R)} {\langle p_{y}| e^{i {\vec q}  {\vec r}}| p_{y}
 \rangle
})+
$$
\begin{equation}
\frac{|c_{\pi}|^{2}}{4}(\cos{(q_{y}R)} {\langle p_{x}| e^{i {\vec q}  {\vec r}}| p_{x}
 \rangle}+
\cos{(q_{x}R)} {\langle p_{y}| e^{i {\vec q}  {\vec r}}| p_{y}
 \rangle
}
),
\end{equation}
$$
	F_{1}(\vec q)=
\frac{|c_{\sigma}|^{2}}{4}(
\cos{(q_{x}R)} {\langle p_{x}| e^{i {\vec q}  {\vec r}}| p_{x}
 \rangle} -
\cos{(q_{y}R)} {\langle p_{y}| e^{i {\vec q}  {\vec r}}| p_{y}
 \rangle}) +
$$
\begin{equation}
\frac{|c_{\pi}|^{2}}{4}(
\cos{(q_{y}R)} {\langle p_{x}| e^{i {\vec q}  {\vec r}}| p_{x}
 \rangle} -
\cos{(q_{x}R)} {\langle p_{y}| e^{i {\vec q}  {\vec r}}| p_{y}
 \rangle
}
).
\end{equation} 
Here $R$ is the $Cu-O$ distance ($R \approx$ 2 \AA) and the matrix elements are
$$
{\langle p_{x,y}| e^{i {\vec q}  {\vec r}}| p_{x,y}
 \rangle} = {\langle j_{0}(qr) \rangle}_{2p}+
$$
\begin{equation}
{\langle j_{2}(qr) \rangle}_{2p}(C^{2}_{0}(\vec q) \mp \sqrt{\frac{3}{2}}(C^{2}_{2}(\vec q)+C^{2}_{-2}(\vec q))),
\end{equation}
$$
{\langle d_{x^{2}-y^{2}}| e^{i {\vec q}  {\vec r}}| d_{x^{2}-y^{2}}
 \rangle} =
{\langle j_{0}(qr) \rangle}_{3d}+
$$
\begin{equation}
\frac{10}{7}{\langle j_{2}(qr) \rangle}_{3d}C^{2}_{0}(\vec q)+
\frac{3}{7}{\langle j_{4}(qr) \rangle}_{3d}(C^{4}_{0}(\vec q)+\sqrt{\frac{35}{2}}(C^{4}_{4}(\vec q)+C^{4}_{-4}(\vec q))). 
\end{equation}
The ${\langle j_{l} \rangle}_{nk}$ in (7-8) is the radial average of the first kind  Bessel function, $C^{l}_{m}$ is the tensor spherical  function.   
The coefficients $c_{d}, c_{p}$ and $c_{\sigma}, c_{\pi}$ obey to the usual relations
\begin{equation}
	|c_{d}|^{2}+|c_{p}|^{2}=1, \qquad |c_{\sigma}|^{2}+|c_{\pi}|^{2}=1.
\end{equation}
These depend on the on site boson number $\hat N$, e.g.
\begin{equation}
|c_{d}|^{2}=|c_{d}^{(0)}|^{2}+(|c_{d}^{(1)}|^{2}-|c_{d}^{(0)}|^{2}){\hat N},
\end{equation}
and have different values for the electron ($N=1$) and hole ($N=0$) centers.

Note that the $b_{1g}$-hole  contribution to the $F_{0}(\vec q)$ coincides formally 
with the form-factor of the $b_{1g}$-hole in  the $CuO_{4}^{6-}$-center of the parent aniferromagnetic matrix
$$
	F(\vec q) = \langle b_{1g} | e^{i {\vec q}{\vec r}}| b_{1g} \rangle =
 |c_{d}|^{2} 
{\langle d_{x^{2}-y^{2}}| e^{i {\vec q}  {\vec r}}| d_{x^{2}-y^{2}}
 \rangle}+
$$
\begin{equation}
 \frac{|c_{p}|^{2}}{2}(\cos{(q_{x}R)} {\langle p_{x}| e^{i {\vec q}  {\vec r}}| p_{x}
 \rangle}+
\cos{(q_{y}R)} {\langle p_{y}| e^{i {\vec q}  {\vec r}}| p_{y}
 \rangle}.
\end{equation}
The non-trivial contribution $F_{1}{\hat \sigma}$ in the form-factor is caused only by the $e_{u}$-hole and has the anomalous   $\vec q$-dependence with the nodes along the $[1,1]$-direction of reciprocal lattice.

In general, the  $F_{0}(\vec q)$,  $F_{1}(\vec q)$ values depend on 
the boson density within the pseudo-JT center and could be considered as the boson number operators
\begin{equation}
F_{0}(\vec q)=F_{0}^{(0)}(\vec q)+F_{0}^{(1)}(\vec q)\,\hat N,
\qquad
	F_{1}(\vec q)=F_{1}^{(0)}(\vec q)+F_{1}^{(1)}(\vec q)\,\hat N.
\end{equation}

In Fig. 3(a-g) we present the $|F(\vec q)|^{2}$ for the $b_{1g}$-hole in a parent compound, the $|F_{0}(\vec q)|^{2}$, $F_{0}(\vec q)F_{1}(\vec q)$ and $|F_{1}(\vec q)|^{2}$ calculated using the known data \cite{clem} on the atomic $Cu3d$ and $O2p$ functions. We pay attention on the anomalous 
$\vec q$-dependence of the $F_{1}(\vec q)$ contribution connected with the local structure (vibronic JT) modes which is responsible for the anisotropic fine structure of the full magnetic form-factor. This structure can be revealed experimentally even for the small $F_{1}(\vec q)$ values.

Finally, the magnetic neutron cross-section in the polar pseudo-JT centers system takes the form
$$
	\frac{d^{2}\sigma}{d \Omega d \varepsilon} \sim
 \sum_{\alpha, \beta = x,y,z}(\delta_{\alpha \beta} - e_{\alpha}e_{\beta})		
 \sum_{m}  \sum_{m'} e^{-i{\vec q}{\vec R}_{m'}}
e^{i{\vec q}{\vec R}_{m}}
\int^{\infty}_{-\infty} dt e^{-i{\omega}t}
$$
$$
\langle (F_{0}^{(0)}(-\vec q)+F_{0}^{(1)}(-\vec q)\,\hat N (m,0)+ 
$$
$$
[F_{1}^{(0)}
(-\vec q)+F_{1}^{(1)}(-\vec q)\,\hat N(m,0)] \sigma^{z}(m,0)) \hat S^{\alpha}(m,0), 
$$
\begin{equation}
(F_{0}^{(0)}(\vec q)+F_{0}^{(1)}(\vec q)\,\hat N(m',t)+ 
$$
$$
[F_{1}^{(0)}(\vec q)+F_{1}^{(1)}(\vec q)\,\hat N(m',t)]\sigma^{z}(m',t))\hat S^{\beta}(m',t) \rangle ,
\end{equation}

where the hybrid "spin-charge-structure" correlation functions are presented. 
The standard approximation for the (13)  implies a breaking apart  the hybrid correlation functions, i.e.
$$
\langle S^{\alpha}(m,0) \sigma^{z}(m,0) \sigma^{z}(m',t) S^{\beta} (m',t) \rangle  	\approx
$$
\begin{equation}
\langle S^{\alpha}(m,0)S^{\beta} (m',t)\rangle  	
\langle \sigma^{z}(m,0) \sigma^{z}(m',t) \rangle  	
\end{equation}
or
$$
\langle S^{\alpha}(m,0) \hat N(m,0) \hat N(m',t) S^{\beta} (m',t) \rangle  	\approx
$$
\begin{equation}
\langle S^{\alpha}(m,0)S^{\beta} (m',t)\rangle  	
\langle \hat N(m,0) \hat N(m',t) \rangle.  	
\end{equation}

The operator structure of the magnetic neutron scattering, an appearance of the hybrid correlation functions in the appropriate cross-section directly points to the possibility to detect the non-magnetic  structure and/or charge dynamic fluctuations by means of the magnetic inelastic neutron scattering. 
This effect displays itself distinctly in the spin subsystem transparency region far from the main spin excitations, when we can neglect the time dependence of the spin correlation function in the relations (14-15). For some cases the complex hybrid structure 
of the magnetic form-factor allows to explain by easily way the observed anomalies in the magnetic inelastic neutron scattering in cuprates.
In particular, it explains an appearance of the strong inter-plane antiferromagnetic like correlations in the "bi-layer" systems such as $YBa_{2}Cu_{3}O_{6+x}$  \cite{Bourges} that are natural for the charge and/or vibronic fluctuations, but improbable  for the pure spin fluctuations at  a weak inter-plane exchange.

\newpage{}

\newpage{}
\begin{center}
	{\bf Figure captions.}
\end{center}

Fig. 1. The illustration of the ground state manifold splitting for the pseudo-JT center.

Fig. 2. The electron density distribution for the $^{1,3}E_{u}$ term of the
$b_{1g}e_{u}(\sigma)$  configuration (left) and the $b_{1g}e_{u}(\pi)$ one (right). 
The dark filling corresponds to the $e_{u}$ orbitals and light filling does to  the $b_{1g}$ ones. The appropriate vibronic states $\sigma_{z}=\pm \frac{1}{2}$ correspond to the $B_{1g}$-type JT mode.

Ðèñ. 3. $\vec q$-dependence (measured in $\AA^{-1}$) of the magnetic form-factor $|F(\vec q)|^{2}$ for $b_{1g}$-hole  in the parent compounds (a); $|F_{0}(\vec q)|^{2}$  at  $c_{\sigma}=1/\sqrt{2}$,  $c_{\pi}=1/\sqrt{2}$ (b) and
$c_{\sigma}=1$,  $c_{\pi}=0$ (c); $F_{0}(\vec q)F_{1}(\vec q)$  at  $c_{\sigma}=1/\sqrt{2}$,  $c_{\pi}=1/\sqrt{2}$ (d) and 
$c_{\sigma}=1$,  $c_{\pi}=0$ (e); $|F_{1}(\vec q)|^{2}$ at $c_{\sigma}=1/\sqrt{2}$,  $c_{\pi}=1/\sqrt{2}$ (f) and
$c_{\sigma}=1$,  $c_{\pi}=0$ (g).    It was taken $c_{d}=0.8$, 
 $c_{p}=0.6$, $q_{z}=0$ for each case. 

}
\end{document}